\documentclass[a4paper,11pt,reqno]{article}
\usepackage{amsfonts}
\usepackage{amsmath}
\usepackage{amssymb}

\usepackage{ tikz, pgfplots, todonotes,ulem}
\usetikzlibrary{decorations.markings,arrows, calc, calligraphy}
\usepackage{caption}
\usepackage[colorlinks=true, pdfstartview=FitV, linkcolor=black, citecolor=black, urlcolor=blue]{hyperref}
\definecolor{shadecolor}{rgb}{0.9, 0.9, 0.81}

\def\red#1{\textcolor[rgb]{0.9, 0, 0}{#1} }

\usepackage{framed,color}
\newlength{\dinwidth}
\newlength{\dinmargin}
\def\C{\mathbb C}
\def\Z{\mathbb Z}

\def\Pcal{{\mathcal P}}
\def\phit{\tilde{\phi}}
\def\varphit{\tilde{\varphi}}
\def\Ht{\tilde{H}}

\def\p{\partial}

\def\g{\gamma}

\def\Qcal{{\mathcal Q}}

\def\be{\begin{equation}}
\def\ee{\end{equation}}
\def\ben{\begin{displaymath}}
\def\een{\end{displaymath}}
\def\baa{\begin{eqnarray}}
\def\eaa{\end{eqnarray}}

\def\ba{\begin{array}}
\def\ea{\end{array}}
\def\la{\label}
\def\p{\partial}
\makeatletter
\@addtoreset{equation}{section}
\makeatother

\newtheorem{definition}{Definition}[section]
\newtheorem{theorem}{Theorem}[section]
\newtheorem{proposition}{Proposition}[section]
\newtheorem{corollary}{Corollary}[section]
\newtheorem{remark}{Remark}[section]
\newtheorem{lemma}{Lemma}[section]

\def\be{\begin{equation}}
\def\ee{\end{equation}}
\def\ben{\begin{displaymath}}
\def\een{\end{displaymath}}
\def\baa{\begin{eqnarray}}
\def\eaa{\end{eqnarray}}

\def\f{\frac}

\def\Acal{{\mathcal A}}
\def\a{\alpha}
\def\b{\beta}
\def\p{\partial}

\def\la{\label}

\def\CC{{\mathcal L}}

\def\Ccal{{\mathcal C}}

\title{Stieltjes-Bethe equations in higher genus and branched coverings with even ramifications}
\author{Dmitry Korotkin}
\begin{document}
\maketitle

{\bf Abstract.} We describe projective structures on a Riemann surface corresponding to monodromy groups which have trivial $SL(2)$ monodromies around singularities and trivial $PSL(2)$ monodromies along homologically non-trivial loops on a Riemann surface. We propose a natural
higher genus analog of Stieltjes-Bethe equations.  Links with branched projective structures and with  Hurwitz spaces with ramifications of even order are established.
We find a higher genus analog of the genus zero Yang-Yang function (the function generating  accessory parameters) and describe its similarity and difference with Bergman tau-function on the Hurwitz spaces.

\tableofcontents
\section{Introduction}

Let $\CC$ be a Riemann surface of genus $g$. 
Consider the scalar differential equation of second order on $\CC$:
\be
(\p^2-U)\varphi=0
\la{eq}
\ee
%\red{Add discussion of $\phi$ versus $\varphi=\phi (d\xi)^{-1/2}$}
where the potential $U$ is allowed to have first and second order poles on $\CC$. 
%The derivative in (\ref{eq}) is taken with respect to some local coordinate $\xi$ on $\CC$. 
The equation (\ref{eq}) is invariant under the choice of the local coordinate if the solution $\varphi$ transforms under a local coordinate change  as a 
$-1/2$-differential while $U$ transforms as $1/2$ of projective connection \cite{HS}. 

In a  local coordinate $\xi$ on $\CC$ one can write $\varphi=\phi(\xi)(d\xi)^{-1/2}$ and $U=u(\xi)(d\xi)^2$; then  (\ref{eq}) takes the
form
\be
\phi_{\xi\xi}-u\phi=0\;.
\la{eq10}
\ee

The ratio $F=\varphit/\varphi$ of two linearly independent solutions  of (\ref{eq}) (called the "developing map") satisfies the 
Schwarzian equation
\be
\{F(\xi),\xi\}=-2u(\xi)
\la{Schw}
\ee
where $\{F,\xi\}=\left(\f{F''}{F'}\right)'-\f{1}{2}\left(\f{F''}{F'}\right)^2$ is the Schwarzian derivative. 
Denote  poles of potential $u$ by $p_1,\dots,p_s$ and consider a set of $2g+s$ generators of the  fundamental group $\pi_1(\CC\setminus\{p_i\}_{i=1}^s, x_0)$ ($x_0$ is a non-singular point of potential $u$) which satisfy the standard relation
\be
\gamma_{p_1}\dots,\gamma_{p_s}\prod_{i=1}^g \a_i\b_i\a_i^{-1}\b_i^{-1} =id\;.
\la{stand}
\ee

The  vector $(\varphit,\varphi)$ of two linearly independent  solutions of (\ref{eq}) which are defined in a neighbourhood of $x_0$ transforms under analytical continuation along any contour $\gamma\in \pi_1(\CC,x_0)$ as follows:
\be
(\varphit,\varphi)\to (\varphit,\varphi) M_\g\;\;, \hskip0.7cm M_\g=\left(\ba{cc} a_\g & b_\g \\ c_\g & d_\g \ea\right)
\ee
where the monodromy matrix  $M_\g$ is a priori defined only up to a sign due to 
spinorial nature of the solutions $(\varphi,\varphit)$.
The corresponding transformation of the solution $F=\varphit/\varphi$ of the Schwarzian equation (\ref{Schw}) looks as follows:
\be
F\to \frac{a_\g F + c_\g}{b_\g F +d_\g}\;;
\ee
these transformations define the $PSL(2,\C)$ monodromy group of the equations (\ref{eq}), (\ref{Schw}).

The lift of this $PSL(2,\C)$ monodromy group to $SL(2,\C)$ (both for holomorphic and meromorphic potentials) 
was discussed in \cite{GKM,BKN,Korot2017}.

There exist several types of monodromy groups which attracted attention over the years:
\begin{itemize}
\item
Trivial $SL(2)$ monodromy groups in genus 0, when for each $\g$ we have $M_\g= I$ (for genus 0, due to existence of a distinguished coordinate $x$, one can talk about $SL(2)$ monodromy representation of (\ref{eq})). The monodromy-free  potentials with second order poles on the Riemann sphere were extensively studied starting from classical works of Heine \cite{He} and Stieltjes \cite{St1,St2,St3}. In terms of a solution $\varphi$ of (\ref{eq}) 
the monodromy-free condition is equivalent to requirement that all residues of $\varphi^{-2}$ vanish; this implies a system of algebraic equations 
for poles and zeros of $\varphi$. Remarkably, this system also appears in the construction of energy eigenstates for quantum $sl(2)$ Gaudin 
system \cite{Gaudin,Sklyanin} where it is known as the system of Bethe equations. Coefficients at $\xi^{-1}$ ($\xi$ is a local parameter)  at poles of $u$ 
are classically known as accessory parameters; in the theory of Gaudin's systems these coefficients correspond to  energy eigenvalues.
When all monodromies are trivial the accessory parameters are given by derivatives of certain "potential" function which can be interpreted as logarithm of discriminant of a rational function $\phi$
 (see for example formula (3.52) of \cite{GW}); this function also carries the name of   "Yang-Yang" function \cite{NRS}. Counting monodromy-free potentials  with fixed poles of $\varphi$ on the Riemann sphere (i.e. counting solutions of Bethe's equations with given poles of $\varphi$) is
 a non-trivial combinatorial problem solvable in terms of Catalan's numbers (see for example \cite{Scherbak2002}). 

\item
Monodromy groups (in any genus) corresponding to so-called "branched projective structures" introduced in  \cite{Man1,Man2} and studied more recently in \cite{Calsam}. Branched projective structures are defined by the condition that all $SL(2)$ monodromies of equation (\ref{eq}) corresponding to generators $\gamma_i$, $i=1,\dots,s$ are  equal to $(-1)^r I$ (where $r$ is the order of the corresponding branch point) while  $PSL(2,\C)$ monodromies along generators $\a_i$ and $\b_i$ 
(monodromies of (\ref{eq}) along $\a_i$ and $\b_i$ are defined only up to a sign) 
form a  representation of
$\pi_1(\CC,x_0)$. In genus zero  branched projective structures with even orders of branching are therefore described by solutions of (\ref{eq}) with trivial $SL(2)$ monodromies.

\end{itemize}

In this paper we consider  solutions of equations (\ref{eq}) and  (\ref{Schw}) on a Riemann surface of an arbitrary genus which correspond to trivial  monodromy groups of equation (\ref{eq}).  "Triviality" here means   that all $SL(2)$ monodromies $M_{\g_j}$
are equal to $I$ while all
 monodromies $M_{\a_j}$, $M_{\b_j}$ (which are defined only up to a sign)
are equal to $\pm I$. We formulate higher genus analogs of Stieltjes-Bethe equations and
introduce two possible definitions of  higher genus  accessory parameters together with their generating functions. 
In analogy to genus zero case the higher genus Stieltjes-Bethe equations define stationary points of 
a generating function in higher genus (this function generalizes the  "Yang-Yang" function of \cite{GW} to higher genus).
The case of genus 1 is treated in detail, closely following the classical  genus zero case.

%We show that monodromy-free potentials (in the above sense) of equation (\ref{eq}) are in 1-to-1 correspondence with 
%points of  Hurwitz spaces with even orders of ramification.

As an intermediate step we discuss branched projective structures corresponding to Abelian monodromy groups where all
monodromy matrices $M_{\a_j}$ and $M_{\b_j}$  are upper-triangular and have unit diagonal. In genus 1 such monodromy groups correspond to solutions of Bethe equations for XYZ Gaudin's model \cite{SklTak96}).

On the other hand, choosing the developing map $F$ as a starting point of construction of monodromy-free potentials on a Riemann surface, we show that such potentials correspond to branched coverings of $\C P^1$ with even orders of all branch points.

To describe these results in more technical terms assume that one of solutions $\varphi$ of equation (\ref{eq}) is a meromorphic 
section of $\chi^{-1}$ where $\chi$ is one of $2^{2g}$ spin line bundles over $\CC$. Moreover, assume that all zeros $x_1\dots x_n$ of $\varphi$ are simple and multiplicities of its poles $y_1,\dots,y_m$ equal to $r_1,\dots,r_m$.
Let us label these points in a universal fashion by introducing the divisor $(\varphi)=\sum_{j=1}^{m+n} d_j p_j$ such that $-2(\varphi)$ is canonical.
Choosing a fundamental polygon $\hat{\CC}$ of $\CC$ in a way compatible with a chosen Torelli marking one can write $\varphi$ as follows in terms of prime-forms (up to a multiplicative constant):
\be
\varphi(x)=  [\Ccal(x)]^{\f{1}{g-1}} e^{\f{2\pi i \langle \beta_1, K^x\rangle}{1-g}}\prod_{j=1}^{m+n} E^{d_j}(x,p_j) 
\la{phiint}\ee
where the prime-forms $E(x,p_j)$ are evaluated at $p_j$ with respect to an arbitrary local coordinate. The holomorphic and 
non-vanishing "multidifferential" $\Ccal(x)$ is given by (\ref{defC}); its properties are discussed in detail in (\cite{Fay92}, p.9-11);
$K^x$ is the vector of Riemann constants with initial point $x$. 
Vectors $\beta_{1,2}\in \Z^g/2$ are defined by equation $-\Acal _{x}((\varphi))+ K^{x} +\Omega \beta_1 +\beta_2=0$
where $\Acal_x$ is the Abel map with initial point $x$ and $\Omega$ is the period matrix of $\CC$ (such vectors exist since the divisor $-2(\varphi)$ is canonical). 

Triviality of $SL(2)$ monodromies $M_{\gamma_j}$ around poles $x_j$  is then equivalent to 
the system of equations
\be
{\rm res}|_{x_j} \f{1}{\varphi^2}=0\;,\hskip0.7cm j=1,\dots,n-1\;.
\la{resint}
\ee

Equations (\ref{resint}) guarantee that the potential $u$ is non-singular at $\{x_j\}$. Using the explicit form (\ref{phiint}) of $\varphi$
one can write equations (\ref{resint}) in a form analogous to genus $0$ Stieltjes-Bethe equations:
\be
\sum_{j=1}^{m} r_j \frac{E'_1(x_k,y_j)}{E(x_k,y_j)} - \sum_{j\neq k, j=1}^{n}  \frac{E'_1(x_k,x_j)}{E(x_k,x_j)}
+\frac{1}{1-g}\frac{\Ccal'(x_k)}{\Ccal(x_k)} +2\pi i \langle\beta_1,\,{\bf v}(x_k)\rangle=0
\la{SBint}
\ee
for $k=1,\dots,n-1$. Equations (\ref{SBint}) are invariant under the change of local coordinates near $\{x_j\}$ and $\{y_j\}$.

Vanishing of  $PSL(2)$ monodromies $M_{\a_j}$, $M_{\b_j}$ is equivalent to
 vanishing of corresponding periods of $\varphi^{-2}$:
 \be
  \int_{a_j}\f{1}{\varphi^2}= \int_{b_j}\f{1}{\varphi^2}=0\;,\hskip0.7cm j=1,\dots,g\;.
  \la{perint}
 \ee
Altogether (\ref{SBint}) and (\ref{perint}), together with the  requirement that divisor $-2(D)$ is canonical   give $3g+n-1$ conditions.  If the Riemann surface $\CC$ remains  fixed the
number of parameters is $m+n$ (positions of $\{y_j\}_{j=1}^m$ and $\{x_j\}_{j=1}^n$); if $\CC$ is allowed to vary we have $3g-3+m+n$ parameters. 
The difference i.e. the dimension of the space of monodromy-free potentials with $m$ poles on Riemann surfaces of genus $g$ equals $m-2$.
This means that generically one
can not fix moduli of $\CC$ and positions of $\{y_j\}$ arbitrarily for $g\geq 1$; this situation is different from the case $g=0$.

Choose a system of local coordinate $\xi_k$ near points of divisor $D$ such that $\xi_k(p_k)=0$.
We denote local coordinates near $y_k$ by $\xi_k$, $k=1,\dots,m$ and near $x_k$ by $\xi_{m+k}$, $k=1,\dots,n$.

 The potential $u$ behaves as follows as
$x\sim y_k$:
\be
u(\xi_{k})\sim \f{r_k(r_k+1)}{\xi_{k}^2} +\f{H_k}{\xi_{k}} + O(1)\;,\hskip0.7cm k=1,\dots,m
\ee
where
\be
H_k=2r_k\left\{\sum_{l\neq k} r_l\f{E'_2(y_l,y_k)}{E(y_l,y_k)} -\sum_{j=1}^n \f{E'_2(x_j,y_k)}{E(x_j,y_k)}\right\}\;,\hskip0.7cm k=1,\dots,m
\la{Hkint}
\ee
and the prime-forms are evaluated at $\{x_k,y_k\}$ with respect to local coordinates $\{\xi_k\}_{k=1}^{m+n}$ ($H_k$ don't depend on the choice of local coordinates near $\{x_k\}$).
As well as in the genus zero case, the higher genus Stieltjes-Bethe  equations (\ref{SBint}) and accessory parameters $H_k$ can be conveniently described in terms of one function: 
\be
\tau_{YY}= e^{-2\pi i \langle \beta_1, \sum_{k=1}^n \Acal_{x_0}(x_k)\rangle}\prod_{k=1}^n \Ccal^{1/(g-1)}(x_k)\prod_{j\neq k,\,j,k=1}^{m+n} E^{d_j d_k}(p_j,p_k)
\la{genint}
\ee
The function $\tau_{YY}$ is considered as a function of $\{x_k\}$, $\{y_k\}$ for a fixed Riemann surface $\CC$.
Namely,  Stieltjes-Bethe equations (\ref{SBint}) are equivalent to conditions
\be
\frac{\p}{\p \xi_{j+m}}\Big|_{\xi_{j+m}=0}\log\tau_{YY}=0\;,\hskip0.7cm j=1,\dots,n-1
\la{YYint}
\ee
while the accessory parameters $H_k$ are given by
\be
H_k=2 \f{\p}{\p \xi_k}\Big|_{\xi_k=0}   \log \tau_{YY}\;,\hskip0.7cm j=1,\dots,m\;.
\la{Htauyy}
\ee
The function $W=2\log \tau_{YY}$ in genus zero is called 
the Yang-Yang function   \cite{NRS,GW}; (\ref{YYint}) is the natural analog of this function in higher genus.

The genus zero version of Stieltjes-Bethe equations arises in construction of energy eigenstates in the XXX Gaudin's model;
the accessory parameters are then equal to eigenvalues of commuting Hamiltonians \cite{Sklyanin}. Equations (\ref{SBint}) in the
genus one case, as well as the corresponding accessory parameters, arise in a similar way in solution of the XYZ Gaudin's model 
\cite{SklTak96}; however, the period conditions (\ref{perint}) don't play an obvious role there.
We are not aware of any existing generalization of these quantum models to higher genus case where higher genus  Stieltjes-Bethe equations (\ref{SBint})  would play a role.  We expect these   models to be quantum versions of $SL(2)$ generalized Hitchin's systems.

Solution of higher genus system (\ref{SBint}), (\ref{perint}) is difficult even in genus zero case. 
However, there is a simple alternative way of constructing the "monodromy-free" potentials of equation (\ref{eq}) whose 
starting point if the developing map $F$. Namely, triviality of monodromies of equation (\ref{eq}) (understood in the above sense)
implies that $F$ is a meromorphic function on $\CC$ with simple poles and branch points of even order:

\begin{theorem}\la{int}
Let  $F$ be a meromorphic function with $n$ simple poles $\{x_j\}_{j=1}^n$  and $m$ critical points
$\{y_k\}_{k=1}^m$
of even multiplicities  $2r_1,\dots,2r_{m}$ on a Riemann surface of genus $g$ i.e. the pair $(\CC,F)$ is an element of the 
Hurwitz space ${\cal H}_g[ 2{\bf r}]$. Then $F$ is a solution of Schwarzian equation with meromorphic potential
\be
u(\xi)=-\f{1}{2}\left\{ F, \xi\right\}\;.
\ee
Moreover, 
$\sqrt{dF}$ is a meromorphic section of a spin line bundle over $\CC$ and
\be
\varphit=\f{F}{\sqrt{dF}}\;,\hskip0.7cm
\varphi=\f{1}{\sqrt{dF}}
\ee
are two linearly independent solutions of (\ref{eq}) which are singular only at $y_1,\dots,y_m$. Corresponding $SL(2)$ monodromies around points $x_k$   as well as $PSL(2)$ monodromies along $(\a_j,\b_j)$ are trivial.
\end{theorem}

This proposition gives a meromorphic potential of (\ref{eq}) starting from an an arbitrary branch covering of given genus with even ramifications. Positions of branch points are given by critical values $z_j=F(y_j)$ of function $F$ i.e. by 
values of $F$ at singularities of potential $U$. Therefore, we can not independently define  moduli
of $\CC$ and positions of singularities of $U$ on $\CC$: changing positions of $z_j$ also changes $\CC$.
This phenomenon is of course related to the fact that in higher genus the total number of equations  (\ref{SBint}), (\ref{perint})
may exceed the number of variables $\{x_j,y_k\}$. In this formulation the problem of counting monodromy-free potentials for fixed $z_1,\dots,z_m$
can be naturally formulated as  (largely unsolved) problem of computation of Hurwitz numbers with even multiplicities of branch points.
We notice also that according to Theorem \ref{int} all monodromy-free potentials on a  Riemann surface of higher genus are 
naturally divided into $2^{2g}$ equivalence classes, depending on spin line bundle defined by divisor $(dF)/2$.

The paper is organized as follows. In Section \ref{gen01} we remind the construction of monodromy-free potentials of (\ref{eq}) in genus 0 and treat the parallel case of genus $1$.
 In Section \ref{triang} we discuss monodromy-free equations (\ref{eq}) on Riemann surfaces of genus $g$,
 and their links with Hurwitz spaces  with branch points of even multiplicity. We
formulate higher genus analog of Bethe-Stieltjes equations  and find generating function of corresponding accessory parameters (the "Yang-Yang function"). 
We also propose an alternative and more invariant way of defining accessory parameters in higher genus case
and discuss their generating function.

\section{Trivial monodromy representations, Stieltjes-Bethe equations and accessory parameters in genera 0 and 1 }
\la{gen01}

\subsection{Genus zero}

We start from reviewing a few facts about monodromy-free equations (\ref{eq}) and (\ref{Schw}) in genus 0.
In this case   one can use the natural coordinate $x$ on $\C$ obtained via stereographic projection of the Riemann sphere.

Assume that the potential $u$ is a rational function of $x$ with poles  of order not higher than 2.
Then triviality of $PSL(2)$ monodromy representation of the  Schwarzian equation 
\be
\{F(x),x\}=-2u(x)
\la{S0}\ee
is equivalent to rationality of the developing map $F(x)$.
In turn, the developing map  defines two linearly independent solutions of the 
linear equation
\be
\phi_{xx}-u(x)\phi=0
\la{eq1}
\ee
via formulas
\be
\phit(x)= \frac{F}{\sqrt{F_x}}\;,\hskip0.7cm \phi= \frac{1}{\sqrt{F_x}}\;.
\la{phi12}
\ee

Rationality of solutions $\phi_{1,2}$ (\ref{phi12}) of the second order linear equation (\ref{eq1}) i.e. triviality of the $SL(2,\C)$ monodromy representation of  (\ref{eq1}) is a stronger requirement than triviality of $PSL(2)$ monodromy of the  Fuchsian equation (\ref{S0}).. Namely, rationality of $\phi_{1,2}$ is equivalent to the condition 
 that  $F_x$ is the square of a rational function i.e. that all poles and zeros of $F_x$ have even order.
In terms of $F$ it means that all critical points of $F$ are of even order and all poles are of odd order.

Assume that poles of $F$ are simple and denote them by $x_1,\dots, x_{n}$. Critical points of $F$ are denoted by $y_1,\dots, y_{m}$ and their multiplicities by $2r_1,\dots,2r_{m}$; we have $n-\sum_{j=1}^{m} r_j=1$.

On Riemann sphere there is unique (up to  a M\"obius transformation) meromorphic spinor with one simple pole; this spinor is given by $(dx)^{1/2}$.
Dividing expressions (\ref{phi12}) by $(dx)^{1/2}$ one gets coordinate-invariant combinations
  $$
  \tilde{\varphi}=\f{\phit}{ \sqrt{dx}}= \f{F}{\sqrt{dF}}\;,\hskip0.7cm \varphi= \f{\phi }{\sqrt{dx}}= \f{1}{\sqrt{dF}}
  $$
   which are meromorphic inverse spinors on $\C P^1$;
\be
\varphi = \frac{\prod_{j=1}^{n} (x-x_j)}{\prod_{j=1}^{m} (x-y_j)^{r_j}} \f{1}{\sqrt{dx}}\;.
\la{phi20}
\ee

Rational functions $F$ with these degrees of critical points and poles form  Hurwitz space which we denote
${\cal H}_0[2{\bf r}]$ where ${\bf r}= (r_1,\dots, r_{m})$.

\begin{remark}\rm
We emphasize the difference between the triviality of the $PSL(2)$ monodromy representation of the Schwarzian equation (\ref{S0}) and the triviality of the $SL(2)$ monodromy representation of the linear equation (\ref{eq1}). Namely, the monodromy representation of (\ref{S0}) is trivial 
iff the developing map $F$ is a rational function. The monodromy representation of  (\ref{eq1}) is trivial only when in addition all points of the divisor 
$(dF)$ are of even order. 
\end{remark}

On the other hand, if the starting point is a rational solution of (\ref{eq1}) of the form (\ref{phi20}) then the rationality of the function $F$ 
(and, therefore, the rationality of the  solution ${\phit}$ (\ref{phi12}))  is equivalent to vanishing of all residues of $\phi^{-2} dx$:
\be
{\rm res}|_{x_j}  \f{dx}{\phi^{2}}=0\;,\hskip0.7cm j=1,\dots, n-1
\la{res0}
\ee
(the residue at $x_n$ vanishes automatically since the sum of residues at all $x_j$ is 0).
Equations (\ref{res0}) are the famous
 Stieltjes-Bethe equations
\be
\sum_{i=1}^{m} \f{r_j}{x_j-y_i}-\sum_{i\neq j,\,i=1}^n \f{1}{x_j-x_i}=0\;,\hskip0.7cm j=1,\dots,n-1\;.
\la{Bethe}
\ee
Equations (\ref{Bethe}) arise in the construction of  eigenstates of Hamiltonians in quantum Gaudin's model \cite{Gaudin,Sklyanin}.

For given critical points $y_j$ and their multiplicities $r_i$ the space of solutions of (\ref{Bethe}) has dimension 1 ($n-1$ equations for $n$ variables $\{x_i\}$). If one  of $x_i$ is fixed  to be, say, $0$ or $\infty$ then the computation of the number of solutions of (\ref{Bethe}) becomes a non-trivial combinatorial problem whose solution is given is terms of
Catalan's numbers (see for example \cite{Scherbak2002}).

A different combinatorial problem arises if one is interested in the number of solutions of the system (\ref{res0}) or (\ref{Bethe}) for
given critical values $z_j= F(y_j)$ of the $n$-sheeted branched covering $z=F(x)$ of genus $0$ defined by function $F$.
This problem is the problem of computation of Hurwitz numbers in genus $0$ for higher order branch points; to the best of our knowledge
it does not have a satisfactory solution except the well-studied case when all branch points are simple except one. %\red{REFERENCE?}

\subsubsection{Accessory parameters and their generating function}

When all zeros $x_j$ of $\phi$ are simple the potential $u$ has at most simple poles at $\{x_j\}$. Moreover, if the linear equation (\ref{eq1}) 
is monodromy-free i.e. the Stieltjes-Bethe equation (\ref{Bethe}) hold, one can easily verify (see for example p.37 of \cite{GW}) that $u$ is non-singular at all $x_j$. Therefore, 
it has the following form:
\be
u(x)=\sum_{j=1}^s \f{r_j(r_j+1)}{(x-y_j)^2}+\sum_{j=1}^s\f{H_j}{x-y_j}
\la{u1}
\ee
(the constant term in this expression is absent since $\phi\sim x $ when $x\to \infty$)
where 
\be
H_j=-2r_j\left(\sum_{k=1}^n \f{1}{y_j-x_k}+\sum_{i=1\,,\; i\neq j}^s \f{r_i}{y_i-y_j}\right)
\la{assp}
\ee
are called {\it accessory parameters}  (see for example \cite{ZT}).  From the point of view of  Gaudin's systems
$H_j$ equal to eigenvalues of (commuting) quantum Hamiltonians corresponding to Bethe eigenstates \cite{Sklyanin}.

Equations (\ref{Bethe}) are equivalent to conditions 
\be
\f{\p \tau_{YY}}{\p x_j} =0
\la{st}
\ee
for function $\tau_{YY}$ given by \cite{SV,GW}:
\be
\tau_{YY}=\frac{\prod_{k<l} (x_k-x_l)\prod_{i<j} (y_i-y_j)^{r_i r_j}}{\prod_{k,j}(x_k-y_j)^{r_j}}
\la{Wdef}
\ee
The function $\tau_{YY}$ is related to function $W$ of \cite{NRS,GW} via 
$\tau_{YY}=e^{W/2}$.
 Introducing the following notation for divisor of $\varphi$:
\be
(\varphi)=\sum_{j=1}^{n+m} d_j p_j = \sum_{j=1}^{n} x_j - \sum_{j=1}^{m } r_j y_j\;,
\ee
the expression (\ref{Wdef})   can be written  in the  "discriminant" form:
\be
\tau_{YY}=\prod_{i<j} (p_i-p_j)^{d_i d_j}\;.
\la{Wshort}
\ee

If $\{x_j\}$ and $\{y_j\}$ are independent  (\ref{Bethe}) then logarithmic derivatives of $\tau_{YY}$ with respect to $y_j$ give the accessory parameters $H_j$ (\ref{assp}): 
\be
2\f{\p }{\p y_j}\log\tau_{YY}=H_j\;;
\la{WH}
\ee
due to (\ref{st}) equations (\ref{WH})  hold also when  $\{x_j\}$ and $\{y_j\}$ are related via Stieltjes-Bethe equations.

 The definition of accessory parameters $H_j$ can be rewritten in coordinate-invariant form as follows:
 \be
 H_j=\f{1}{2}{\rm res}|_{y_j}\frac{S_{B}-S_{dF}}{dx}
 \la{defHj}
 \ee
 where $S_{dF}=-2U$ is the (meromorphic) projective connection given by
 $S_{dF}(\xi)= \{F(\xi),\xi\} (d\xi)^2= -2u(x)(dx)^2$ and  $S_B$ is the projective connection given by Schwarzian derivative of coordinate $x$ with respect to any other local coordinate $\xi$ on $\C P^1$: $S_B=\{x(\xi),\xi\} (d\xi)^2$. 
 
 The projective connection $S_B$ (\ref{Ber0}) is nothing but (up to a factor $1/6$) the constant term in expansion of the bidifferential
 \be
 B(x,y)=\frac{dx dy}{(x-y)^2}
 \la{Bergm0}
 \ee
  near diagonal $x=y$ in a local parameter $\xi$. Namely,
 \be
 S_B(x)=\frac{1}{6}\left[\frac{dx  dy}{(x-y)^2}- \frac{d\xi(x) d\xi(y)}{(\xi(x)-\xi(y))^2}\right]\Big|_{x=y}\;.
 \la{Ber0}
 \ee
 The numerator in (\ref{defHj}) is a difference of two projective connections i.e. a  meromorphic quadratic differential;
  dividing it by $dx$ we get an abelian differential with   well-defined  residue.
  The bidifferential $B(x,y)$ (as well as its generalization to higher genus), carries the name of {\it canonical bidifferential}
  \cite{Fay73}, and is also called the {\it Bergman bidifferential} \cite{Tyurin}; the corresponding projective connection $S_B$ 
 is thus called the {\it Bergman projective connection}. In coordinate $x$ on $\C P^1$  obviously $S_B(x)\equiv 0$; thus $x$ is the developing map of projective structure given by projective connection $S_B$.
 
 The definition (\ref{defHj}) of accessory parameters as it stands does not admit a  generalization to genus greater than 1
 (in genus one case one has a natural choice of global coordinate, thus a straightforward generalization of (\ref{defHj}) does make sense).
In higher genus one can replace $x$ either by one of uniformization coordinates (Fuchsian, Schottky) or by developing map of some chosen projective structure.  For example, a natural choice would be to take the developing map corresponding to higher genus Bergman projective connection (which, however, depends on Torelli marking of $\CC$).

\subsubsection{An alternative definition of accessory parameters and Bergman tau-function}

Equations (\ref{defHj}) can be modified to give an alternative  universal definition of accessory parameters which can be generalized to  any genus. In this alternative definition one replaces $dx$ in the denominator by another natural 1-form,  $dF$;  then the definition of accessory parameters looks as follows:
 \be
 \Ht_j=\f{1}{2}{\rm res}|_{y_j}\frac{S_B-S_{dF}}{d F}\;.
 \la{defHt}
 \ee 
 
 The generating function for accessory parameters $\Ht_j$ {\it with respect to critical values} $z_j$ is also known; it is given by $\log(\tau_B^3)$ where $\tau_B$ is the  so-called {\it Bergman tau-function} \cite{IMRN}. 
 Namely,
 \be
 \f{\p }{\p z_j} \log (\tau_B^3)= \Ht_j\;.
 \la{genHt}
 \ee
The expression for $\tau_B^3$ looks as follows (this expression can be deduced from formulas  of \cite{IMRN,IMRN1} via simple computation):
 \be
 \tau_B^3=\prod_{j<l} (p_i-p_j)^{\frac{2d_i d_j (d_i+d_j+1)}{(2d_i+1)(2d_j+1)}}\;.
 \la{tauB0}
 \ee

  The equation (\ref{genHt}), as well as the alternative definition (\ref{defHt}) of accessory parameters, admits a unique natural generalization to
 any genus.

Notice the similarity and difference between expressions (\ref{Wshort}) and (\ref{tauB0}): both expressions are singular only along hyperplanes $p_i=p_j$ but
with different degrees.

\subsubsection{Parabolic triangular $SL(2)$  monodromies in genus 0}

Let equation (\ref{eq1}) possess one solution $\phi$ of the form (\ref{phi20}) but don't assume the  Stieltjes-Bethe  equations (\ref{Bethe}) to hold.
Then the "second" solution 
\be
\phit=\phi\int^x \frac{dx}{\phi^2}
\la{phi1tri}
\ee
is non-singlevalued on $\C P^1$: it has logarithmic singularities at $\{x_i\}_{i=1}^{n}$. 

In this case the potential $u=\phi''/\phi$ generically has poles at all zeros  $\{x_j\}$ and poles $\{y_j\}$ of $\phi$. 
Since integrals of $\phi^{-2}dx$ around $\{y_j\}$ vanish, all $SL(2)$ monodromy matrices of equation (\ref{eq1}) around 
all $\{y_i\}$ are trivial:  $M_{y_i}=I$. 
On the other hand, monodromies around $x_j$ are triangular matrices of the form 
\be
M_{x_j}= \left(\ba{cc} 1 & 0 \\ \Pcal_{x_j} & 1 \ea\right)\;\; \hskip0.5cm  \in SL(2,\C)
\la{trm}
\ee
with $\Pcal_{x_j}=2\pi i\, {\rm res}|_{x_j} \phi^{-2}dx$.
The $SL(2)$ monodromy group generated by   matrices (\ref{trm}) is  Abelian.

\subsection{Trivial $SL(2)$ monodromies in genus one}

The genus one case can be treated in parallel to the genus zero case due to existence of a natural coordinate on $\CC$. Let $\CC$ be an elliptic curve with periods $1$ and $\sigma$ (corresponding to some Torelli marking)
and with flat coordinate $x$ such that $v=dx$ is the normalized holomorphic differential on $\CC$.
Then triviality of all $PSL(2)$ monodromies of the Schwarz equation (\ref{Schw}) is equivalent to the developing map $F$ is a meromorphic function on $\CC$.

On the other hand, triviality of $SL(2)$ monodromies $M_{\gamma_i}$ of the linear equation (\ref{eq}) around singularities of $U$ is a stronger requirement: it implies that,  as in genus zero case,  all points of divisor $(dF)$ have even multiplicity.
Then formulas (\ref{phi12}) again give two linearly independent solutions of equation (\ref{eq}) with trivial $SL(2)$ monodromies around poles of $U$.

The (unique up to a constant) holomorphic spinor on $\CC$ is given by $\sqrt{dx}$; it is a holomorphic section of the 
only odd spin line bundle  (this line bundle corresponds to characteristics $[1/2,1/2]$ irrespectively of Torelli marking; see the discussion of correspondence between theta-characteristics and spin line bundles in the next section which is  devoted to the higher genus case).

  Assume that one of solutions of equation $(\p^2 -U)\varphi=0$ is such that $\varphi^{-1}$
is a  meromorphic section of the odd spin line bundle
and all of  zeros of $\varphi$ are 
simple.

 Positions of poles in $x$-coordinate  will be denoted by $y_1,\dots,y_{m}$ and positions of zeros by $x_1,\dots,x_{n}$; let
$(\varphi)=\sum_{i=1}^{n} x_i-\sum_{i=1}^{m} r_i y_i$ and $r_1+\dots+r_{m}=n$. 
Since divisor $-2(\varphi)$ is canonical  (that is the same as trivial for $g=1$) and divisor $-(\varphi)$ corresponds to the odd spin line bundle one can introduce $\beta_{1,2}\in \Z$ such that 
\be
-\sum_{i=1}^{n} x_i+\sum_{i=1}^{m} r_i y_i+\beta_1\sigma+\beta_2=0 \;.
\la{choicefd}
\ee
%(we add $1/2$ to $\beta_{1,2}$ to keep correspondence with higher genus: in $g=1$ the vector of Riemann constants equals $(\sigma+1)/2$).  
Then solution $\varphi$ takes the form
\be
\varphi= \phi(x)(dx)^{-1/2}
\la{phi2ell}
\ee
where 
\be
\phi(x) = \f{\prod_{j=1}^{n} \theta_1(x-x_j)}{\prod_{j=1}^{m} \theta_1^{r_j}(x-y_j)} e^{2\pi i  \beta_1 x} 
\la{phigen1}\ee
and  $\theta_1(x)=\theta[1/2,1/2](x,\sigma)$. 

Using periodicity properties of $\theta_1$ it is easy to verify, taking (\ref{choicefd}) into account, that $\varphi$ is  a single-valued meromorphic function on $\CC$.

The second linearly independent solution $\phit(x)$ of equation $\phi_{xx}-u\phi=0$ is as usual given by 
\be
\phit(x)=\phi\int_{x_0}^x \f{dx}{\phi^{-2} }\;.
\la{phi1ell}
\ee
Due to  uniqueness of the spinor $(dx)^{1/2}$  which is used to get function $\phi$ from the inverse spinor $\varphi$ in a unique way one can  define an $SL(2)$ monodromy representation of (\ref{eq1}) in genus 1 unambigously as monodromy of pair $(\phit,\phi)$.
Namely, the monodromy representation 1 is generated by matrices $M_\a, M_\b, M_{x_1},\dots,M_{x_{n}}$ with generators of the fundamental group chosen such that 
\be
M_{\b} M_{\a} M_{\b}^{-1} M_\a^{-1} M_{x_n},\dots,M_{x_1}=I\;.
\ee
For any $\gamma\in\pi_1(\CC\setminus\{x_j\}_{j=1}^{n})$ the monodromy matrix $M_\g$ is again given by
\be
M_\g= \left(\ba{cc} 1 & 0 \\ \Pcal_{\gamma} & 1 \ea\right)\;\; \hskip0.5cm  \in SL(2,\C)
\la{trm}\ee
with $\Pcal_{\gamma}=\oint_{\gamma} \phi^{-2} dx$.
We therefore get the following 
\begin{proposition}
The pair of solutions $(\phit,\phi)$ of equation 
\be
\phi_{xx}-u\phi =0
\ee
given by (\ref{phigen1}), (\ref{phi1ell}), has all trivial $SL(2)$ monodromies on elliptic curve $\CC$ iff
the following equations hold:
\be
{\rm res}|_{x_j} \f{dx}{\phi^2}=0\;,\hskip0.7cm j=1,\dots n-1\;,
\la{resell}
\ee
\be
\int_{a}\f{dx}{\phi^2}=\int_b\f{dx}{\phi^2} =0\;.
\la{perell}
\ee
\end{proposition}

Together with condition (\ref{choicefd}) that sum of zeros of $\varphi$ equal to the sum of its poles up to an integer combination of periods, the    equations   (\ref{resell}) and (\ref{perell}) give $n+2$ equations. The number of parameters $\{x_i,y_i\}$ is 
$n+m-1$ ($-1$ is due to translational invariance on the torus). This  gives $m-3$ for the dimension of the moduli space of monodromy-free 
potentials on a fixed elliptic curve with fixed numbers of poles and zeros of $\phi$. If the elliptic curve is allowed to vary this dimension becomes $m-2$.
Therefore, for given $m$ and $\{r_i\}_{i=1}^{m}$ the monodromy-free configurations of poles $\{y_i\}$ and zeros $\{x_i\}$ of $\phi$ exist   for $m\geq 2$.
On the other hand, if the  curve $\CC$ and positions of poles $y_1,\dots,y_{m}$ are fixed, the monodromy-free condition gives 
$n+2$ equations for $n$ zeros $\{x_i\}$. These equations therefore are solvable only for special configurations of $(\sigma,y_1,\dots,y_{m})$.

In analogy to the genus zero case, one can alternatively start the description of monodromy-free potentials from 
a meromorphic developing map $F$ having  $n$ simple poles and 
$m$ critical points $y_1,\dots,y_{m}$ of multiplicities $2r_1,\dots,2r_{m}$ ($\sum_{i=1}^{m} r_i=n$). Corresponding solutions $\phit$ and $\phi$ are then given by
\be
\phit=\frac{F}{\sqrt{F_x}}\;,\hskip0.7cm 
\phi=\frac{F}{\sqrt{F_x}}
\ee
which are both meromorphic (i.e. the corresponding $SL(2)$ monodromy representation is trivial)  if $\sqrt{F_x}$ is also a meromorphic function.
Denote  critical values of $F$ by $z_j=F(y_j)$. The dimension of the moduli space of monodromy-free potentials (allowing 
the module of $\CC$ to vary) gives the same number $m-2$: a simultaneous shift and simultaneous rescaling of all critical values
correspond to shift and rescaling of the developing map $F$; this does not change the potential $u$.
% \red{Why an arbitrary M\"obius ransf of $F$ is not allowed? that would give $m-3$.}

 The natural counting problem in this setting is to find the number of inequivalent pairs $(\CC,F)$ with given $z_1,\dots,z_{m}$.
This number equals the Hurwitz number $h_{1,n}(2r_1,\dots,2r_{m})$; as well as in the genus zero case such problem does not have a 
satisfactory solution by now. Notice that in this counting problem the module of the Riemann surface is not fixed.

\subsubsection{Stieltjes-Bethe equations on elliptic curve and accessory parameters}

 Equations (\ref{resell}) are direct elliptic analogs of genus zero Stieltjes-Bethe equations (\ref{Bethe}); they can be explicitly written as follows:
 \be
 \sum_{i=1}^{m} r_i\f{\theta_1'(x_j-y_i)}{\theta_1(x_j-y_i)} -\sum_{i\neq j,\,i=1}^n\f{\theta_1'(x_j-x_i)}{\theta_1(x_j-x_i)}=0\;,\hskip0.7cm
 j=1,\dots,n-1\;.
 \la{SBell}
 \ee

 Similarly to the genus 0 case, equations (\ref{SBell}) are equivalent to equations (\ref{st}):
 \be
 \f{\p }{\p x_j}\log\tau_{YY}=0
 \la{stell}\ee
   with
  \be
 \tau_{YY}=\frac{\prod_{k<l} \theta_1(x_k-x_l)\prod_{i<j} \theta_1^{r_i r_j}(y_i-y_j)}{\prod_{k,j}\theta_1^{r_j}(x_k-y_j)}
\ee
 or, equivalently,
 \be
 \tau_{YY}= \prod_{i<j} \theta_1^{d_i d_j}(p_i-p_j) \;.
 \la{Well}
 \ee

 Equations (\ref{SBell}) again guarantee that the potential $u$ is non-singular at $\{x_i\}$. Analyzing the singularity structure 
 of $u$ near $y_i$ we get, analogously to (\ref{u1}) that the quadratic residue of $U(x)= u(x) (dx)^2$ at $y_k$  equals to $r_k(r_k+1)$; therefore,
 \be
 u(x)=\sum_{k=1}^{m}\left[ -r_k(r_k+1) f'(x-y_k) + H_k f(x-y_k)\right] +C
 \la{uell}
 \ee
 where 
 $$f(x)=\f{\theta_1'(x)}{\theta_1(x)}$$ 
 and
  the accessory parameters $H_k$ are given by 
  \be
  H_k=2 r_k\left\{ \sum_{l\neq k}r_l f(y_k-y_l) -\sum_{j=1}^n f(y_k-x_j)\right\}\;.
  \la{accell}
  \ee
 The constant $C$ is not equal to $0$, in contrast to genus zero case; it can be expressed in terms of $\{x_j,y_j\}$ but we don't need this expression here.

 The formulas (\ref{uell}), (\ref{accell})  are obtained by expressing $u=\phi''/\phi= (\phi'/\phi)'+ (\phi'/\phi)^2$ 
 and taking into account that
 \be
 \f{\phi'}{\phi}= \sum_{j=1}^n f(x-x_j)-\sum_{k=1}^{m} r_k f(x-y_k)
 \ee
 to give
 $$
 u(x)=\sum_{k=1}^{m} \left(r_k f'(x-y_k)+ r_k^2 f^2(x-y_k)\right)
 $$
 $$
 +2\sum_{k=1}^{m} r_k f(x-y_k)\left[\sum_{l\neq k} r_l  f(x-y_l)-\sum_{j=1}^n f(x-x_j)\right]+ 
 \{ {\rm terms \;\;non-singular\;\;at}\;\;\{y_k\}\}
 $$
  Analyzing the singular parts of $u$ at $y_j$ we come to (\ref{uell}), (\ref{accell}).

  Accessory parameters $H_j$ can be expressed via the developing map $F$ as in 
  (\ref{defHj})
  \be
 H_j=\f{1}{2}{\rm res}|_{y_j}\frac{S_{B}-S_{dF}}{dx}
 \la{defHjel}
 \ee
  where $x$ is the flat coordinate on the torus, and the Bergman projective connection is given by the formula (27) and the last formula on p.34 of \cite{Fay73}:
  $S_B=-2\f{\theta_1'''}{\theta_1'}(0)(dx)^2$. Although $S_B$ does not vanish in coordinate $x$ (as in genus zero case),
  it is holomorphic and, therefore,   does not contribute to the residue in (\ref{defHjel}).

 \begin{proposition}
The accessory parameters $H_k$ are given by logarithmic derivatives of function $\tau_{YY}$  defined by (\ref{Well}), with respect to $y_k$:
 \be
 H_k=2\f{\p }{\p y_k}\log\tau_{YY}
 \la{Hkell}
 \ee
 in both cases: when $(x_j,y_j)$ are independent (i.e. related only by condition that divisor $-2(\varphi)$ is trivial) 
 and when the Stieltjes-Bethe equations (\ref{SBell}) hold. 
 %\red{Check: what happens with periods of $\varphi^{-2}$? These are also coordinates..} 
 The Riemann surface $\CC$ remains fixed under differentiation.
 \end{proposition}
 {\it Proof.} The proof is obtained by direct calculation. 
 
 $\Box$

 \begin{remark}\rm
  Elliptic Stieltjes-Bethe equations (\ref{SBell}) arise in construction of energy eigenstates  in the $XYZ$ (elliptic) Gaudin model (see (5.15) of \cite{SklTak96} for the case $\nu=0$), similarly to  appearance of the genus zero Stieltjes-Bethe equations (\ref{Bethe}) in the ordinary XXX Gaudin model.
 Accessory parameters (\ref{accell}) coincide with eigenvalues of commuting Hamiltonians 
 in the elliptic Gaudin's model ($H_k$ coincide with residues of  $\tau dx$ where $\tau$ is the function given by (5.17) of \cite{SklTak96}).
On the other hand, conditions of vanishing of $a$- and $b$- periods of $\varphi^{-2}$ (\ref{perell})
 which provide triviality of monodromies $M_a$ and $M_b$, do not arise in this context. 
 
 %\red{Discuss the link of SB equations for XYZ Gaudin model with  even branch projective structures in genus 1???}
 \end{remark}

  As well as in genus zero case, the numerator of (\ref{defHjel}) is well-defined in any genus, while the 
  1-form $dx$ in the denominator does not admit a natiural generalization. On the other hand, replacing 
  $dx$ by $dF$ one can define alternative accessory parameters $\tilde{H}_j$ via formula (\ref{defHt}). 
  The generating function of $\tilde{H}_j$ with respect to critical values of function $F$ 
  (i.e. with respect to branch points of the branched covering defined by function $F$) is the 3rd power of the Bergman tau-function $\tau_B$ 
  as in (\ref{genHt}):
  \be
  \f{\p \log\tau_B}{\p z_k}= \f{1}{3} \tilde{H}_k\;.
  \ee
   An explicit formula for $\tau_B$ is given in Prop. 3 of \cite{IMRN}.
   
 %\red{ check!
 % \be
  % \tau_B^3=\prod_{j<k}\theta_1(p_j-p_k)^{\frac{2d_i d_j (d_i+d_j+1)}{(2d_i+1)(2d_j+1)}}\;.
 %\la{tauB1}\ee}
%  where as before $\sum_{i=1}^{n+m} d_i p_i=\sum_{j=1}^n x_j - \sum_{j=1}^{m} r_j y_j$.
  
  % Expression (\ref{tauB1}) is a straightforward generalization of the expression (\ref{tauB0}) in genus zero.

\section{Parabolic triangular and trivial monodromy groups in higher genus}
\la{triang}

Let now $g\geq 2$. This case is different from $g=0$ or $g=1$ by the following reason. In genus zero there exists a unique 
(up to a M\"obius transformation) meromorphic spinor $\sqrt{dx}$. This spinor is used to rewrite the second order equation 
$(\partial^2-U)\varphi=0$ using $x$ as independent variable and $\phi=\varphi\sqrt{dx}$ as dependent variable in the coordinate form $\phi_{xx}+u\phi=0$. In turn, this allows to define the $SL(2)$ monodromy representation. In genus one the transformation from $\varphi$ to $\phi$ is performed using the flat coordinate $x$ on elliptic curve and the only holomorphic odd spinor $\sqrt{dx}$.

For higher genus one can choose the universal coordinate in various ways (say, use the Fuchsian or Schottky uniformization coordinate).
Another ambiguity is in the choice of a special spinor used in transformation from $\varphi$ to $\phi$ (one natural way to define such spinor is proposed  in \cite{BKN}).

%since on a Riemann surface $\CC$ one can not find a special section of one of spin line bundles. 
Let us assume that there exists a pair $(\varphit,\varphi)$ of solutions of (\ref{eq}) such that all monodromy matrices  have the form
\be
M_\g= \pm\left(\ba{cc} 1 & 0 \\ * & 1 \ea\right)\;\; \hskip0.5cm  \in PSL(2,\C)\;.
\la{trm}
\ee
Then  $\varphi$ is a meromorphic section of a line bundle $\chi^{-1}$ where $\chi$ is (one of $2^{2g}$) spin  line bundles
 over $\C$ ( $\chi^2$ is the canonical line bundle); thus $deg(\varphi)=1-g$.

Consider the divisor $D$ of $\varphi$. Assuming that all zeros of $\varphi$ are simple we have
\be
D=\sum_{j=1}^{m+n} d_j p_j=\sum_{j=1}^{n}  x_j - \sum_{j=1}^{m } r_j y_j \;;
\la{div}
\ee
since the divisor $-2D$ is canonical,
the divisor $-D$ corresponds to  some (even or odd) spin line bundle $\chi$ and
$n-\sum_{j=1}^{m} r_j=1-g\;.$

Introduce some system of local coordinates $\{\xi_j\}$ near $\{p_j\}_{j=1}^{n+m}$ such that $\xi_1,\dots,\xi_m$ are local coordinates near $y_1,\dots,y_m$, respectively while $\xi_{m+1},\dots,\xi_{m+n}$ are local coordinates near $x_1,\dots,x_n$, respectively.

% \red{LOCAL COORDINATES $\xi_j$ HERE!}

The corresponding potential  $u$ has second order poles at points $y_j$ 
of the following form:
\be
u(\zeta_j)=\f{r_j(r_j+1)}{\xi_j^2}+ \f{H_j}{\xi_j}+O(1)\;, \hskip0.7cm j=1,\dots,m
\la{asyj}
\ee
 While the quadratic residue $r_j(r_j+1)$ is invariant under the change of local coordinate $\xi_j$
the coefficients $H_j$ (the "accessory parameters") are not.

At the points $x_j$ the potential has in general poles of first order. 
The condition of holomorphy of $U$ at $x_j$ is given by the following lemma:
\begin{lemma}
The potential $U$ is non-singular at $x_j$ iff ${\rm res}|_{x_j}\varphi^{-2}=0$.
\end{lemma} 
The {\it proof} is an elementary local computation: since $\phi$ has a first order zero at $x_j$ it  can be written in a local 
coordinate $\xi\equiv\xi_{m+j}$
as follows: $\phi= a\xi(1+b\xi+\dots)$. Then $\f{\phi_{\xi\xi}}{\phi}=2b \xi^{-1}+ O(1)$ while $\phi^{-2}= \xi^{-2}-b\xi^{-1} + O(1)$ which implies the 
statement of the lemma.

$\Box$

\subsection{Parabolic triangular monodromies}

%Let the genus $g$ of $\CC$ be greater or equal than 2.

Choose some canonical basis of cycles (the Torelli marking) $\{a_i,b_i\}_{i=1}^g$ in $H_1(\CC,\Z)$. The dual basis
of holomorphic differentials $v_1,\dots,v_g$ is normalized by $\int_{a_i}v_j=\delta_{ij}$ and the period matrix of $\CC$ is defined by
$\Omega_{ij}=\int_{b_i}v_j$. Introduce the theta-function $\Theta(z)=\Theta(z,\Omega)$ for  $z\in \C^g$. Choose a fundamental $4g$-gon 
$\tilde{\CC}$ of $\CC$ in a way which agrees with the choice of canonical cycles $(a_i,b_i)$ (in fact, the contours forming the sides of the fundamental polygon 
correspond to a set of generators $(\a_j,\b_j)$ of the fundamental group of $\CC$).
Denote vector of Riemann constants corresponding to a point $x_0\in\CC$ by $K^{x_0}$ and denote the Abel map corresponding to 
initial point $x_0$ by $\Acal_{x_0}(x)$: $(\Acal_{x_0})_{j}(x)=\int_{x_0}^x v_j$. 

Introduce the prime-form $E(x,y)$ \cite{Mum} that is a multi-valued $(-1/2,-1/2)$-form on $\CC$. 
Holonomies of $E(x,y)$ with respect to, say, the first variable $x$ along cycle $a_j$ equal $1$ while along cycle $b_j$ equal
\be
e^{-\pi i \Omega_{ii}-2\pi i  (\Acal(x)-\Acal(y))}
\la{holE}\ee

For any point $p_i\in D$ we shall use the notation
$$
E(x,p_i)=E(x,y)\sqrt{d \xi_i(y)}\Big|_{y=p_i}\;.
$$

A half-integer characteristic  corresponding to divisor $D$ (\ref{div}) is defined by equation:
\be
-\Acal _{x_0}(D)+ K^{x_0} +\Omega \beta_1 +\beta_2=0
\la{defbeta}
\ee
where $\beta_{1,2}\in (\Z/2)^g$ are vectors with integer or half-integer components; we denote $[\beta]=[\beta_1,\beta_2]$.
Integration contours in (\ref{defbeta}) are chosen inside of the fundamental polygon $\tilde{\CC}$.
Under a change of Torelli marking by a symplectic transformation  and the corresponding change of the fundamental polygon the   vectors $\beta_{1,2}$ 
are transformed according to formulas given on page 8 of \cite{Fay73}. For a contemporary discussion of the theory of  theta-characteristics see \cite{Farkas}.
% \red{they give the Arf -invariant of the symplectic group}.

Let us also introduce the following multi-valued $g(1-g)/2$-differential on $\CC$ (see (1.17) of \cite{Fay92}):
\be
\Ccal(x)=\f{1}{W[v_1,\dots,v_g](x)}\sum_{i_1+\dots+i_g=g} \frac{\partial^g \Theta(z)}{\p^{i_1} z_1,\dots,\p^{i_g} z_g}\Big|_{z=K^x}
v_{i_1}(x)\dots v_{i_g}(x)
\la{defC}
\ee
where 
\be
W[v_1,\dots,v_g](x)={\rm det}_{1\leq i,j\leq g} ||v_j^{(i-1}(x)||
\ee
 is the Wronskian determinant of basic holomorphic differentials. The differential $\Ccal(x)$ is non-singular and non-vanishing on $\CC$.
Holonomies of $\Ccal(x)$ along cycles $a_j$ and $b_j$ are equal to $1$ and 
\be
e^{-\pi i (g-1)^2\Omega_{jj}-2\pi i  (g-1) K_j^x}\;,
\la{holC}\ee
respectively.

\begin{proposition}
 Let a divisor $D$ (\ref{div}) be such that  $-2D$ is canonical. Define the characteristic vectors with $\Z/2$ components by (\ref{defbeta}) and denote by $\chi$ the spin line bundle corresponding to divisor $-D$.
 Then a section $\varphi$ of the line bundle $\chi^{-1}$ such that $(\varphi)=D$ is given by
\be
\varphi(x)=  [\Ccal(x)]^{\f{1}{g-1}} e^{\f{2\pi i \langle \beta_1, K^x\rangle}{1-g}}\prod_{j=1}^{m+n} E^{d_j}(x,p_j) \;.
\la{phi2}
\ee
%Choosing some   $\phi(z)=\varphi \sqrt{dz}$ on the opposite sides of the fundamental polygon are related by
%\be
%\phi(x+ a_j)= e^{2\pi i (\beta_1)_j}\phi(x)\;,\hskip0.7cm
%\phi(x+ b_j)= e^{2\pi i (\beta_2)_j}\phi(x)\;.
%\la{holsp}
%\ee}
\end{proposition}
{\it Proof.}  The tensor weight of expression (\ref{phi2})  to the sum of the contribution from $\Ccal$ (which equals $\f{1}{g-1}\f{g(1-g)}{2}=-\f{g}{2}$) and contribution $\f{g-1}{2}$ from the product of prime-forms (which equals to the product of ${\rm deg} D =(1-g)$ and $-1/2$). Thus the total tensor weight of $\phi$ equals $-1/2$.

The product of prime-forms gives poles and zeros of necessary degree at  divisor $D$ while other factors are non-singular and non-vanishing. 
It remains to verify that $\varphi^{-2}$ is a well-defined meromorphic differential on $\CC$.

The only term in (\ref{phi2})  which produces a non-trivial multiplier along cycle $a_j$ is the exponential term. The
transformation of the exponential term follows from holonomy of the vector of Riemann constants along $a_j$ given by
 (see for example \cite{Mum})
\be
K^{x+ a_j} = K^x+(g-1)e_j\;,
\la{holKa}\ee
where $e_j=(0,\dots,1,\dots,0)^t$ with $1$ standing at $j$th place. Thus the holonomy of $\varphi$ along cycle $a_j$  equals 
$ e^{2\pi i (\beta_1)_j}=\pm 1$; however, the sign of this holonomy can not be given an invariant meaning independent of the choice of fundamental polygon. What is however important for us in that $\varphi^{-2}$ has trivial monodromy along all $a$-cycles.

The holonomy of $\varphi^{-2}$ along cycle $b_j$ can be computed multiplying the holonomies (\ref{holE}) of the prime-forms
with holonomy (\ref{holC}) of $\Ccal(x)$ and taking into account the definition of vectors $\beta_{1,2}$ (\ref{defbeta}). Another relation which needs to be used in this computation is the holonomy of the vector of Riemann constants along cycle $b_j$:
\be
K^{x+ b_j} = K^x+(g-1)\Omega e_j\;.
\la{holKb}\ee

Then the holonomy of $\varphi$  along $b_j$  is  given by the exponent of the following expression:
\be
-\pi i (g-1)\Omega_{jj} -2\pi i K_j^x -2\pi i \langle\beta_1,\;\Omega e_j\rangle
-\sum_{k=1}^{s_1+m} \left(\pi i d_k \Omega_{jj}+2\pi i d_k((\Acal_{x_0})_j(x)- (\Acal_{x_0})_j(p_k))\right)
\la{cal1}
\ee
where the first two terms originate from $\Ccal(x)$, the third term gives the holonomy of the exponential term in (\ref{phi2}), and the sum gives holonomy of the product of prime-forms. Furthermore, using the relations  $\sum_{k=1}^{m+n} d_k=1-g$ and (\ref{defbeta}) we can rewrite the sum over $k$ in (\ref{cal1}) as follows
\be
\pi i \Omega_{jj}+ 2\pi i (g-1) \Acal_{x_0}(x) +2\pi i (K^{x_0}_j +(\Omega\beta_1)_j+(\beta_2)_j)\;,
\la{cal2}
\ee
which, after substitution into (\ref{cal1}), gives
\be
2\pi i\left[ -K^x_j+K_j^{x_0}+ (g-1)\Acal_{x_0}(x) - \langle\beta_1,\,\Omega e_j\rangle +(\Omega\beta_1)_j +(\beta_2)_j\right]\;.
\la{cal3}
\ee
The first three terms in (\ref{cal3}) vanish due to the transformation property of the vector of Riemann constants. The terms containing $\beta_1$ vanish due to symmetry of matrix $\Omega$. 
Therefore, the holonomy factor of $\varphi$ along cycle $b_j$ formally equals
$e^{2\pi i (\beta_2)_j}$. We therefore conclude that $\varphi^{-2}$ is a meromorphic differential on $\CC$ which proves the proposition:
the choice of the square root of the canonical bundle $\chi$ whose section is given by $\varphi^{-1}$ is determined by the choice of divisor $D$.

$\Box$

Choosing some local coordinate $\xi$ on $\CC$ we define the following function $h(\xi)$ in the corresponding coordinate chart:
\be
h(\xi)=\f{d}{d\xi}\log\left(\varphi(x) \sqrt{d\xi(x)}\right)\;.
\la{h}
\ee
Then the potential $U$ can be written in this local coordinate as follows:
\be
U(\xi)= (h_{\xi}+ h^2)(d\xi)^2\;.
\la{uh}
\ee

\begin{proposition}
Define $-1/2$-differential $\varphi$ on $\CC$ via (\ref{phi2}) and introduce $\varphit$ via
\be
\varphit(x)=\varphi(x)\int_{x_0}^x \varphi^{-2}
\la{defphi1}
\ee
for some base point $x_0$. Then $\varphit$ and $\varphi$ form a pair of linearly independent solutions of equation (\ref{eq}) with potential $U$ given by (\ref{uh}).

The corresponding $PSL(2)$ monodromy representation looks as follows: for any $\gamma\in \pi_1(\CC\setminus \{x_j\}_{j=1}^{n})$ the monodromy matrix is given by
\be
M_\gamma=\pm\left(\ba{cc} 1 & 0\\ \int_\gamma\varphi^{-2} & 1\ea\right)
\la{montr}
\ee
\end{proposition}

The {\it proof} is straightforward.  Since $\varphi^{-1}$ is a meromorphic spinor on $\CC$
the only source of a non-trivial $PSL(2)$ monodromy is the non-singlevaluedness of the Abelian integral $\int_{x_0}^x \varphi^{-2}$; the contributions to monodromy matrices are given by periods of $\varphi^{-2}$ in the lower off-diagonal term as in (\ref{montr}). 

$\Box$

Notice that the off-diagonal terms of monodromy matrices around points $x_j$ are given (up to the factor of $2\pi i$) by residues of $\varphi^{-2}$ at points $x_j$.  Moreover, although $PSL(2)$ monodromies around $y_j$ are trivial, the $SL(2)$ monodromies of equation (\ref{eq}) around $y_j$ are well-defined and are equal either to $I$ or $-I$, depending on parity of $r_j$.

If we denote the Abelian differential $\varphi^{-2}$ by $W(x)$ then the Schwarz equation corresponding to these solutions of (\ref{eq}) looks as follows:
\be
\left\{\int_{x_0}^x W,\, \xi(x)\right\} =-2 u (\xi(x)) 
\ee
for any local parameter $\xi(x)$.

\subsection{Projective structures with even branching}

The previous construction gives examples of so-called branched projective structures  \cite{Man1,Man2}. For recent developments in this area we refer to \cite{Calsam,GKM}. The branched projective structures  on a Riemann surface  can be characterized in terms of equations (\ref{eq}) with meromorphic potential in the following way: {\it all $PSL(2)$ monodromies of the developing map around singularities of $U$ are trivial}.
The  branched projective structures arising in our context are described in the following definition:
\begin{definition}
A branched projective structure is called even if orders of all branch points are even. Equivalently, $SL(2)$  monodromies of equation (\ref{eq}) around singularities of
potential $U$  are trivial i.e. all solutions  of (\ref{eq}) are single-valued in small discs around singularities of $U$. 
\end{definition}

%Another reformulation of the evenness of branch projective structure is that  at any singularity $x_0$ of $u$ the most singular term looks as
%\be
%u(\xi)=\frac{r_0(r_0+1)}{\xi^2}+\dots
%\ee
%where $\xi$ is a local parameter near $x_0$ and $r_0\in \Z$
%(for an arbitrary branched projective structure the 
%coefficients $r_0$ can be either integer or half-integer).

% Examples of even branched projective structures naturally from our previous construction.

\begin{proposition}
Let a Riemann surface $\CC$ and a divisor $D$ (\ref{div}) (such that $-2D$ is canonical), satisfy the following additional conditions
\be
{\rm res}\big|_{x_j} \frac{1}{\varphi^2}=0\;,\hskip0.7cm j=1,\dots,n
\la{rescon}
\ee
where $\varphi$ is given by (\ref{phi2})
(conditions (\ref{rescon}) are equivalent to the requirement that $\varphi^{-2}$ is a meromorphic abelian differential of second kind). Then the Abelian integral $F(x)=\varphit/\varphi=\int_{x_0}^x\varphi^{-2}$ 
gives a developing map of  an even branched projective structure. 
%(i.e.  monodromies of all solutions of equation (\ref{eq}) around poles of $u$ are trivial). 
Monodromy matrices along remaining loops $(\a_i,\beta_i)$  corresponding to such branched projective structures have the triangular parabolic form (\ref{montr}) i.e. these matrices are determined by $a$- and $b$- periods of the differential $\varphi^{-2}$.
\end{proposition}

Although this proposition is essentially tautological, the condition of vanishing of residues of $\varphi^{-2}$  is highly non-trivial for $n>1$ since all  poles $x_j$ of $\varphi^{-2}$
are of even order. The cases $n=0$ (when $\varphi^{-2}$ is a holomorphic differential with even orders of its zeros) and $n=1$ (when $\varphi^{-2}$ has only one pole of order 2
and zeros of even order zeros of $\varphi^{-2}$) are therefore special. 

\begin{proposition}
Let in divisor $D$ (\ref{div}) either $n=0$ or $n=1$. Then $F(x)=\int_{x_0}^x \varphi^{-2}$ where $\varphi$ is given by (\ref{phi2}) is the developing map of an even branched
projective structure on $\CC$ with branch points at $y_1,\dots,y_{m}$. 
The order of branching at $y_j$ equals $2r_j$. 
\end{proposition} 

To prove this proposition it is sufficient to notice that for $n=0,1$ conditions (\ref{rescon}) are empty. 

Moreover, for $n=0$ the starting point of the construction can be any holomorphic abelian differential (say, $W$) on $\CC$ with zeros of even orders $2r_1,\dots,2r_{m}$. Then $\sqrt{W}$ is a  spinor on $\CC$ (i.e. a section of one of $2^{2g}$ square roots of canonical  bundle). One can define the solution $\varphi$ via $\varphi=1/\sqrt{W}$ and the developing map 
via the abelian integral of $W$ (branched projective structures of this type were discussed in \cite{Calsam}). 
We notice that such branched projective structures, as well as all other projective structures discussed in this paper, can be divided into $2^{2g}$ equivalence classes labelled by spin structure (in particular, they can be naturally divided into the even and odd ones).

\subsubsection{Stieltjes-Bethe equations in higher genus and even branched projective structures}

For  $n\geq 2$ conditions (\ref{rescon}) are non-trivial. 
\begin{proposition}
Let $g\geq 2$, $m\geq 2$.  Then the triviality of $SL(2)$ monodromies around all points of divisor $D$ is
equivalent to the following system of $n-1$ equations:
\be
\sum_{j=1}^{m} r_j \frac{E'_1(x_k,y_j)}{E(x_k,y_j)} - \sum_{j\neq k, j=1}^{n}  \frac{E'_1(x_k,x_j)}{E(x_k,x_j)}
+\frac{1}{1-g}\frac{\Ccal'(x_k)}{\Ccal(x_k)} +2\pi i \langle\beta_1,\,{\bf v}(x_k)\rangle=0
\la{SB}
\ee
for $k=1,\dots,n-1$, where ${\bf v}$ is the vector $(v_1,\dots,v_g)^t$ of normalized holomorphic 1-forms. 
Notation $v_j(x_k)$ 
is used for $\frac{v_j(x)}{d\xi_{m+k}(x)}$ where $\xi_{m+k}(x)$ is a local parameter near $x_k$. The same convention is used to define $\Ccal(x_k)$. Derivatives  $\Ccal'(x_k)$ are taken with respect to the same local parameters. 
For any two points of divisor $D$ we define
\be
E(p_j,p_k)=E(x,y)\sqrt{d\xi_j(x)}\sqrt{d\xi_k(y)}\Big|_{x=p_j,y=p_k}
\la{Epp}
\ee
 Index $1$ denotes derivative with respect to the first argument of $E$.
The system (\ref{SB}) does not depend on the choice of local coordinates near $x_k$ and $y_j$.
\end{proposition}
{\it Proof.} Equations (\ref{SB}) are obtained by computing residues (\ref{rescon}) at points $x_k$ which are second order poles of the 1-form $\varphi^{-2}$. Independence of equations (\ref{SB}) on the choice of local coordinates near $x_j$ and $y_k$ follows from invariance of residue conditions (\ref{rescon}).

$\Box$

\begin{definition}
Equations (\ref{SB}) are called  {\bf higher genus Stieltjes-Bethe equations.}
\end{definition}
 
% In genus zero case they were first formulated by Stieltjes \cite{St1,St2,St3} and more recently appeared in construction of energy eigenstates in quantum Gaudin's model 
%\cite{Gaudin,Sklyanin}.

The dimension of the subspace of the moduli space of branched projective structures arising via this construction equals $2g-2+m$. This dimension is equal to the number of parameters ($3g-3$ moduli of $\CC$ plus $n+m$ points of divisor $D$) minus the number of conditions 
($g$ conditions stating that divisor $-2D$ is canonical and $n-1$ equations (\ref{SB})). We recall that the dimension of the 
moduli space of all branched projective structures (with $m$ branch points and fixed branching orders) equals  $6g-6+m$.

Notice also that the space of branch projective structures obtained via our construction consists of  $2^{2g}$ components
labelled by  square roots of canonical line bundle.

\subsubsection{Accessory parameters and their generating function}

Accessory parameters $H_j$ in higher genus are defined by the local behaviour (\ref{asyj}) of potential $u$ near $y_j$.
For solution $\varphi$ given by (\ref{phi2}) they are given by

\be
H_k=2r_k\left\{\sum_{l\neq k} r_l\f{E'_2(y_l,y_k)}{E(y_l,y_k)} -\sum_{j=1}^n \f{E'_2(x_j,y_k)}{E(x_j,y_k)}\right\}
\la{Hkdefh}
\ee
(the index $2$ denotes the derivative of the prime-form with respect to its second variable)
which looks as  a direct generalization of the  of the low genus formulas (\ref{assp}) and (\ref{accell}).

Notice that the definition (\ref{Hkdefh}) of accessory parameters in more ambiguous than in genera 0 and 1
where a distinguished coordinate on $\CC$ exists.
In $g\geq 2$  accessory parameters $H_k$  depend on the choice of local coordinates near 
all $\{x_j\}$ and $\{y_j\}$.

Similarly to genera 0 and 1, both the higher genus Stieltjes-Bethe equations (\ref{SB}) and accessory parameters (\ref{Hkdefh})
can be described in terms of a single scalar function.  
\begin{definition}
The "Yang-Yang function"    $\tau_{YY}$ for $g\geq 2$ is defined by:
\be
\tau_{YY}= e^{-2\pi i \langle \beta_1, \sum_{k=1}^n \Acal_{x_0}(x_k)\rangle}\prod_{k=1}^n \Ccal^{1/(g-1)}(x_k)\prod_{j\neq k} E^{d_j d_k}(p_j,p_k)
\la{eW1}
\ee
where the prime-forms are evaluated with respect to the local coordinates $\{\xi_j\}$:
\be
E(p_j,p_k)= E(x,y) \sqrt{d\xi_j(x)}\sqrt{d\xi_k(y)}\Big|_{x=p_j,\,y=p_k}\;.
\la{defE}
\ee
\end{definition}

Equivalently (recall that $\sum_{i=1}^{n+m} d_i p_i= \sum_{j=1}^n x_j-\sum_{k=1}^{m} r_k y_k$), the formula (\ref{eW1}) can be written as follows:
\be
\tau_{YY}= e^{-2\pi i \langle \beta_1, \Acal_{x_0}(x_k)\rangle}\prod_{k=1}^n \Ccal^{1/(g-1)}(x_k)\frac{\left\{\prod_{j\neq k,\,j,k=1}^n E(x_j,x_k)\right\}\left\{\prod_{j\neq k,\,j,k=1}^{m} E^{r_j r_k} (y_j,y_k)\right\}}{\prod_{k=1}^n \prod_{j=1}^{m} E^{r_j} (x_k, y_j)}\;.
\la{eW2}
\ee

Comparing the formula for $\tau_{YY}$ (\ref{eW2}) with the higher genus Stieltjes-Bethe equations (\ref{SB}) and
the expressions (\ref{Hkdefh}) for accessory parameters $H_j$ we get the following
\begin{proposition}
The higher genus Stieltjes-Bethe equations (\ref{SB})  are equivalent to vanishing of derivatives of function $\tau_{YY}$ (\ref{eW2})
with respect to zeros $x_j$ which in terms of corresponding local parameters $\xi_k$ can be written as follows:
\be
\f{\p }{\p \xi_k}\Big|_{\xi_k=0}\log\tau_{YY}=0\;,\hskip0.7cm k=m+1,\dots,m+n
\la{stath}
\ee
where the Riemann surface $\CC$ is assumed to be fixed.  The system of equations (\ref{stath}) does not depend on the choice of
local coordinate $\xi_{m+1},\dots,\xi_{m+n}$ near $x_1,\dots,x_n$.

Accessory parameters (\ref{Hkdefh}) are given by
\be
H_k=2 \frac{\p}{\p \xi_k}\Big|_{\zeta_k=0} \log\tau_{YY}\;, \hskip0.7cm k=1,\dots,m\;.
\la{Hjhg}
\ee
\end{proposition}
%\red{check $\xi$ versus $\zeta$ everywhere}

\begin{remark} \rm
We are not aware of existence of a quantum model whose Bethe equations  coincide with conditions (\ref{SB})
of triviality of $SL(2)$ monodromies around $y_j$. However, we expect that such system (presumably a quantum version of a
generalized Hitchin system) should exist, in analogy to genera 0 and 1.

\end{remark}

The definition (\ref{Hkdefh}) of accessory parameters $H_j$ can be made less ambiguous if one uses one of special coordinates
to write the expansion (\ref{asyj}) of the potential neat points $y_j$ (say, Schottky or Fuchsian uniformization coordinates).
Another choice of such coordinate would be to introduce the developing map $\zeta_B(x)$ of equation $(\partial^2-\f{1}{2}S_B)\varphi=0$, where $S_B$ is the Bergman projective connection in higher genus. The definition of $S_B$ is discussed in detail in 
\cite{Fay73,Tyurin,JDG,BKN}: similarly to (\ref{Ber0}), $S_B$ is, up to a factor $1/6$, a constant term in the expansion of the
canonical bimeromorphic differential $B(x,y)=d_x d_y\log E(x,y)$ near the diagonal $x=y$; $S_B$ depends on Torelli marking of $\CC$.

The developing map $\xi_B$ solves the Schwarzian equation $\{\xi_B(x),\cdot\}= S_B(x)$ (we don't know what monodromy 
representation corresponds to this equation).
Denoting the developing map of equation (\ref{eq}) with trivial monodromies by $F$ (which is a meromorphic function with simple poles at $\{x_j\}$ and critical points at $y_j$) we can write the definition (\ref{asyj}), (\ref{Hkdefh}) of accessory parameters $H_j$ of (\ref{eq}) which 
corresponds to coordinate $\xi_B$ in the following form:
\be
H_j^B=\f{1}{2}{\rm res}|_{x=y_j}\frac{S_B-S_{dF}}{d\xi_B(x)}\;.
\la{BerAcc}
\ee
If the function $\tau_{YY}$ (\ref{eW1}) is also computed using the local coordinate $\xi_B$ near all points of divisor $(\varphi)$
then the accessory parameters $H_j^B$ are obtained as logarithmic derivatives of $\tau_{YY}$ with respect to $\xi_B(y_j)$ as in (\ref{Hjhg}).

\subsection{Trivial monodromies and Hurwitz spaces with even ramifications}

A special case of  branched projective structures  is the case when the developing map has all trivial $PSL(2,\C)$  monodromies i.e. the developing map is a meromorphic function on the Riemann surface.
If, moreover, such branched projective structure is even then all zeros and poles of the differential $dF$ should be of even degree such that all
{\bf $SL(2)$} monodromies around $y_j$ are equal to $I$.

Then both solutions, $\varphi$ and $\varphit$ of (\ref{eq}) are sections of $\chi^{-1}$ for some spin line bundle $\chi$ over $\CC$.  This implies vanishing of all periods of the meromorphic differential $\varphi^{-2}$, and not only residues at $x_j$ i.e. we get the following
\begin{proposition}
Let $\varphi$ be a section of line bundle $\chi^{-1}$ given by (\ref{phi2}) where divisor $D$ is defined by (\ref{div}).
Let, moreover,  period of 1-form $\varphi^{-2}$ over any cycle $s\in H_1(\CC\setminus\{x_j\}_{j=1}^{n}$ vanish, i.e.
\be
\int_{a_i} \varphi^{-2}=\int_{b_j} \varphi^{-2} =0\;,\hskip0.7cm i=1,\dots, g\;,
\la{SB1}
\ee
\be
{\rm res}\big|_{x_j} \varphi^{-2}=0\;,\hskip0.7cm j=1,\dots,n-1\;.
\la{SB2}
\ee
Then all $SL(2)$ monodromies of equation (\ref{eq}) with potential $U(\xi(x))=-\frac{1}{2}\left\{\int^x \varphi^{-2},\xi(x)\right\}(d\xi)^2$,
around singularities of $U$ are trivial. All $PSL(2)$ monodromies along generators $\a_j$ and $\b_j$ are 
also trivial.

The dimension of the space $S_g[ {\bf r}]$ of the inverse spinors (\ref{phi2}) corresponding to given ${\bf r}=(r_1,\dots, r_{m})$ on Riemann surfaces of given genus is given by 
\be
{\rm dim}\,S_g[ {\bf r}]= m-2\;;
\la{dimH}
\ee
the space $S_g[ {\bf r}]$  is naturally stratified into $2^{2g}$ strata corresponding to different spin line bundles over $\CC$.
%\red{This dimension coincides also with the dimension of the space of monodromy-free potentials with $m$ poles of the form 
%(\ref{asyj}).}
\end{proposition}
{\it Proof.} The number of parameters in (\ref{phi2}) equals $3g-3+n+m-g=2g-3+n+m$ (which is the sum of the dimension of moduli space of Riemann surfaces
and number  of points of divisor $D$ such that divisor $-2D$ is canonical). The number of conditions in (\ref{SB1}), (\ref{SB2}) equals
$2g+n-1$. The difference of these two numbers equals $m-2$, as stated.

$\Box$

%Notice that the dimension of the moduli space of monodromy-free potentials $u$ equals $m-3$ i.e. it differs from (\ref{dimH}) by 1.
%This difference is due to the fact that there is a one-parametric freedom (up to a trivial rescaling) in the  choice of the "second" solution $\phi$; one could equivalently start from any $\tilde{\phi}_2= \phi+ \alpha\phit$ for $\alpha\in \C$ without changing the potential $u$.

If  $\varphi$ satisfies conditions (\ref{SB1}), (\ref{SB2}) 
then the developing map i.e. the Abelian integral 
\be
F(x)=\int_{x_0}^x \varphi^{-2}
\la{defF}\ee
for an arbitrarily chosen base point $x_0$
is single-valued on $\CC$ i.e. it
is in fact  a meromorphic function on $\CC$ with simple poles at $x_1,\dots,x_{n}$ 
and critical points at $y_1,\dots, y_{m}$ (of multiplicities $2r_1,\dots,2r_{m}$, respectively).

The developing map $F$ defines an $n$-sheeted covering 
\be
z=F(x)
\la{cover}
\ee
of Riemann sphere with coordinate $z$; the critical (branch) points of function $F$ are  $y_1,\dots,y_{m}$ 
of multiplicities $2r_1,\dots,2r_{m}$, respectively, while all poles are simple.

The space of functions $F$ with such branching and pole  profile is the Hurwitz space ${\cal H}_g[2{\bf r}]$. 
If the coverings corresponding to functions $F$ and $\alpha F+\beta$ are identified for any $\a,\b\in \C$ (such coverings correspond to the same $\varphi$, up to a multiplicative constant), we get the quotient   
${\cal H}_g[2{\bf r}]/\{F\sim \a F+\b\}$
which is naturally identified with the space $S_g[ {\bf r}]$ of monodromy-free potentials with double  poles  with
quadratic residues  given by (\ref{asyj}) .

The natural coordinates on the space ${\cal H}_g[2{\bf r}]$ are the critical values of function $F$:
\be
z_j=F(y_j)\;,\hskip0.7cm j=1,\dots,m
\la{branch}
\ee
By transformation of the form $z\to \a z+\b$ one can always put $z_1=0$, $z_2=1$. Then the remaining $m-2$ critical values $z_3,\dots,z_{m}$ can be used as local coordinates on the space of monodromy-free  potentials.

The above discussion can be summarized as follows.

\begin{proposition}\la{Hurprop}
Let  $F$ be a meromorphic function with $n$ simple poles  and $m$ critical values
of even multiplicities  $2r_1,\dots,2r_{m}$ on a Riemann surface of genus $g$ i.e.
$(\CC,F)\in {\cal H}_g[2{\bf r}]$.Then $F$ is a solution of Schwarzian equation with meromorphic potential
\be
u(\xi)=-\f{1}{2}\left\{ F, \xi\right\}\;;
\ee
$\sqrt{dF}$ is a meromorphic section of a spin line bundle over $\CC$, the equation  (\ref{eq}) has all trivial $SL(2)$ monodromies
and two linearly independent solutions of (\ref{eq}) are given by
\be
\varphit=\f{F}{\sqrt{dF}}\;,\hskip0.7cm
\varphi=\f{1}{\sqrt{dF}}\;.
\ee
\end{proposition}

Notice that in Prop.\ref{Hurprop} no restriction on genus is necessary, it is also valid in $g=0$ and $g=1$ cases.

\subsubsection{Alternative definition of accessory parameters and Bergman tau-function on Hurwitz spaces}

In the Hurwitz picture one can propose an alternative definition of accessory parameters (denoted by $\tilde{H}_j$) following the definition (\ref{genHt}) in genera 0 and 1:
\be
 \Ht_j=\f{1}{2}{\rm res}|_{y_j}\frac{S_B-S_{dF}}{d F}
 \la{defHt1}
\ee
The parameters $ \Ht_j$ are generated by the Bergman tau-function on Hurwitz space with respect to  critical values $z_j$:
\be
 2\f{\p }{\p z_j} \log (\tau_B^{3/2})= \Ht_j
 \la{genHt1}
 \ee
where $\tau_B^{3/2}$ is given by expression which resembles (\ref{eW1}) \cite{IMRN1}: 
\be
\tau_B^{3/2}= \Qcal^{\f{g-1}{2}}e^{-\f{\pi i}{4} \langle\beta_1,\Omega \beta_1\rangle}\prod_{j\neq k} \tilde{E}^{d_j d_k}(p_j,p_k)
\la{taubany}\ee
where
\be
\Qcal = \sqrt{dF(x)}[\Ccal(x)]^{\f{1}{1-g}}  e^{\f{2\pi i \langle \beta_1,K^x\rangle}{g-1}} \prod_{j=1}^{m+n} \tilde{E}^{-d_j}(x,p_j)
\la{Qcal}\ee
is an $x$-independent factor. All prime-forms in (\ref{taubany}) and (\ref{Qcal}) are evaluated at points of $y_j$ with respect to the so-called {\it distinguished local coordinates} 
\be
\zeta_j(x)=\left[F(x)-F(y_j)\right]^{1/(2r_j+1)}
\la{disty}
\ee
Near $x_j$ the distinguished local coordinates are given by $\zeta_{m+j}(x)=1/F(x)$. 

Namely,
\be
\tilde{E}(x,p_j)= E(x,y)\sqrt{d\zeta_j(y)}\Big|_{y=p_j}\;,\hskip0.7cm
\tilde{E}(p_j,p_k)= E(x,y)\sqrt{d\zeta_j(x)}\sqrt{d\zeta_k(y)}\Big|_{x=p_j,\,y=p_k}
\la{defEt}
\ee

This makes the product of prime-forms in (\ref{taubany}) 
significantly different from the product of prime-forms in (\ref{eW1}) which are evaluated in an arbitrary system of local coordinates near $\{p_j\}$
(specifying these coordinates to be given by the "Bergman" coordinate $\xi_B$ one gets accessory parameters in the form (\ref{BerAcc})).

The difference in the evaluation of prime-forms (\ref{defEt}) in (\ref{taubany}) and (\ref{defE}) in (\ref{eW1}) implies, in particular, different asymptotics near the boundary, when two points of the divisor $D$ coalesce. Namely, denoting by $t$ a coordinate near the boundary of moduli space which corresponds to merging of $p_j$ and $p_k$ the function $\tau_{YY}$ behaves in the limit $t\to 0$, similarly to the genus zero case (\ref{Wdef}) as 
$$
\tau_{YY}\sim t^{d_j d_k}(1+O(t))\;.
$$
The asymptotics of $\tau_B^3$, on the other hand, looks as follows as $t\to 0$ (similarly to genus 0 formula (\ref{tauB0})):
\be
\tau_B^3 \sim t^{\f{2d_k d_j (d_k+d_j+1)}{(2d_j+1)(2d_k+1)}} (1+ O(t))
\ee

Notice also that variation with respect to $z_j$  in (\ref{genHt1}) changes the conformal structure of the Riemann surface $\CC$, in contrast to variation with respect to positions of $y_j$ in formulas (\ref{Hjhg}) for $H_j$.

Finally, we would like to mention an open question of computing the "Yang-Yang" function in a different context,
when the monodromy representation of (\ref{eq}) is generic \cite{NRS}. Then the Yang-Yang function is defined as
 generating function between complex Fenchel-Nielsen coordinates on an $SL(2)$ character variety and canonical 
 coordinates on $T^*{\mathcal M}_{g,n}$. In this formulation the problem of computation of $\tau_{YY}$ is more complicated;
 a partial result towards description of this function was obtained in \cite{BKN} where it was shown that $\tau_{YY}$ transforms as a
 section of Hodge line bundle over ${\mathcal M}_{g,n}$ under a change of Torelli marking.

{\bf Acknowledgements.}  I thank G.Kappagantula, C.Norton, V.Tarasov and M.Shapiro for helpful discussions.
This   was supported in part by the Natural Sciences and Engineering Research Council of Canada grant
RGPIN/3827-2015, FQRNT grant "Matrices Al\'eatoires, Processus Stochastiques et Syst\`emes Int\'egrables" (2013--PR--166790) and by Alexander von Humboldt Stiftung. 
The author thanks   Max-Planck Institute for Gravitational Physics in Golm (Albert Einstein Institute) for  hospitality during the preparation of this work.

\end{document}